# Magnetic Field Effects on the Structure of Radiatively Cooling Protostellar Jets

E.M. de Gouveia Dal Pino & A.H. Cerqueira

*University of São Paulo, Instituto Astronômico e Geofísico, Brasil*

## Introduction

Magnetic fields have been neglected in most of the analytical and numerical modeling of the structure of protostellar jets since the inferred estimates of their intensity (B ~$10^{-6}$ - $10^{-5}$ G) suggest that they may be not dynamically important along the flow. However, after amplification by compression behind the shocks at the jet head, they may become relevant as they are carried backward with the shocked gas to fill the cocoon that envelopes the jet.

Considerable amount of work on magnetized, adiabatic, light jets has been done to study extragalactic jets (see, e.g., [1] for a review). Lately, these MHD studies have been extended to heavy, adiabatic jets [11,12, 13], however, the general effects of **B**-fields on the global evolution and morphology of radiatively cooling jets, have just started to be explored in numerical studies [2, 3, 9, 10, 12]. In the present work, with the help of fully 3-D smooth particle magnetohydrodynamical (SPMHD) simulations, we extend these prior investigations by exploring the nonlinear effects of magnetic fields (in ~ equipartition with the gas) in the structure of *radiatively cooling, overdense* jets and test their importance on the dynamics of protostellar jets. A detail analysis of the results presented here may be found in [3].

We solve the MHD equations in the ideal approximation using a modified version of the purely hydrodynamical SPH code developed by de Gouveia Dal Pino & Benz [5] [see also 4, 6-9]. In the computational domain, the ambient gas is represented by a 3-D rectangular box filled with particles initially distributed in a cubic lattice array. The jet of radius $R_j$ is injected into the bottom of the box and propagates through the ambient medium up to a distance of ~30 $R_j$. In the transverse directions (y and z), the box has dimensions ~24 $R_j$. We use outflow conditions at the boundaries. The particles are smoothed out by a spherically symmetric kernel function of width h (initially, h=0.4 $R_j$ and 0.2 $R_j$ for the ambient and the jet particles). The radiative cooling (due to collisional excitation and recombination) is implicitly calculated using a time-independent cooling function for a gas of solar abundances cooling from T~$10^6$ to $10^4$ K. Two initial magnetic field configurations are assumed: an uniform longitudinal magnetic field permeating both the jet and the ambient medium [B = ($B_o$,0,0)]; and a helical magnetic field, both of which extending to the ambient medium. In the latter, the maximum strength corresponds to the longitudinal component at the jet axis; the azimuthal component attains a maximum value ($B_\phi = 0.39 B_o$) at ~3 $R_j$.

The initial parameters of the simulations depicted below (chosen to resemble the conditions found in protostellar jets) are: $\eta = n_j/n_a = 3$ (the number density ratio between jet and ambient medium); $M_a = v_j/c_a = 24$ (the ambient Mach number with a jet velocity $v_j = 400$ km/s); $M_{ms,j} = 28$ (the jet magnetosonic number); $\beta = p_{th}/p_B = 1$ (the thermal to the magnetic pressure ratio at the jet axis); and $q_{bs} = d_{cool,bs}/R_j = 8$ (the ratio of the radiative cooling length in the postshock gas behind the bow shock to the jet radius; the correspoding ratio behind the jet shock is $q_{js} = 0.3$).

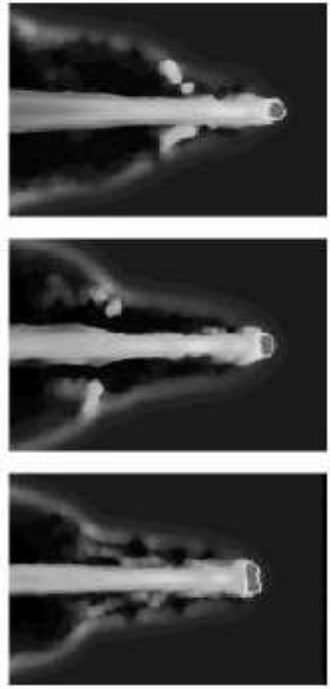

**Figure 1**: Gray-scale representation of the midplane density of the head of radiative cooling jets with **η=3 and $M_a$=24**: a hydrodynamical jet (**HD3r, top**), an MHD jet with initial longitudinal **B**-field (**ML3r, middle**), and an MHD jet with initial helicoidal **B**-field (**MH3r, bottom**) after they have propagated ~30$R_j$ (t/$t_d$ = 1.65, $t_d$ = $R_j/c_a$ =38 yr, where $c_a$ is the ambient sound speed). The maximum density attained at the shell is $n_{sh}/n_a$ =210, 153, and 160 (from top to bottom).

## Results and Conclusions

We find that magnetic fields have important effects on the dynamics and structure of radiative cooling jets, especially at the head. As an example, Fig. 1 depicts the midplane density section of the head of three supermagnetosonic, radiatively cooling jets with $\eta=3$ after they have propagated ~30 $R_j$. The time evolution of the corresponding velocity and magnetic field distributions are presented in Figs. 2, 3. The pure HD jet (HD3r, Figs. 1, 2, top) develops a cold dense shell at the head, due to the cooling of the shock-heated jet material. It becomes Rayleigh-Taylor (R-T) and thermal unstable and breaks into blobs that spill into the cocoon [e.g, 5] and show a resemblance with the Herbig-Haro objects which are observed at the head of protostellar jets. These cold blobs also develop in the MHD jet with longitudinal field (ML3r, Figs. 1, 2, 3, middle), but in this case they detach from the beam as they are expelled backward to the cocoon. In the MHD jet with helical magnetic field (MH3r, Figs.1, 2, 3, bottom), initially a cold shell also develops, but the toroidal B-field component, which is amplified by compression in the shocks at the head, increases the cooling length behind the jet shock and reduces the density enhancement. As a consequence, the shell tends to be stabilized against R-T instability and the clump formation is thus inhibited.. These results suggest that a predominantly helical field configuration is unlikely at the jet head of protostellar jets.

Both longitudinal and helical magnetic field geometries improve jet collimation. In both magnetic configurations. The associated magnetic pressures (~$B^2/8\pi$) cause an increase in the total pressure of the cocoon, relative to the pure HD case, which collimate the beam and excites the (the fastest growing) small-wavelength pinch modes of the MHD K-H instability. These modes over-confine the beam and drive approximately equally spaced internal shocks (Figs.2, 3). We note also, the appearance of an MHD helical mode which causes some jet twisting close to the ML3r head. However, the weakness of the induced shocks makes it doubtful that they could produce by themselves the bright knots observed in the highly overdense, radiatively cooling protostellar jets.

Finally, the beam pinching detected in the MHD cooling jets discussed above is also found in adiabatic jets with similar initial conditions. As expected [e.g., 4, 12], they are more intense and numerous in the adiabatic case.

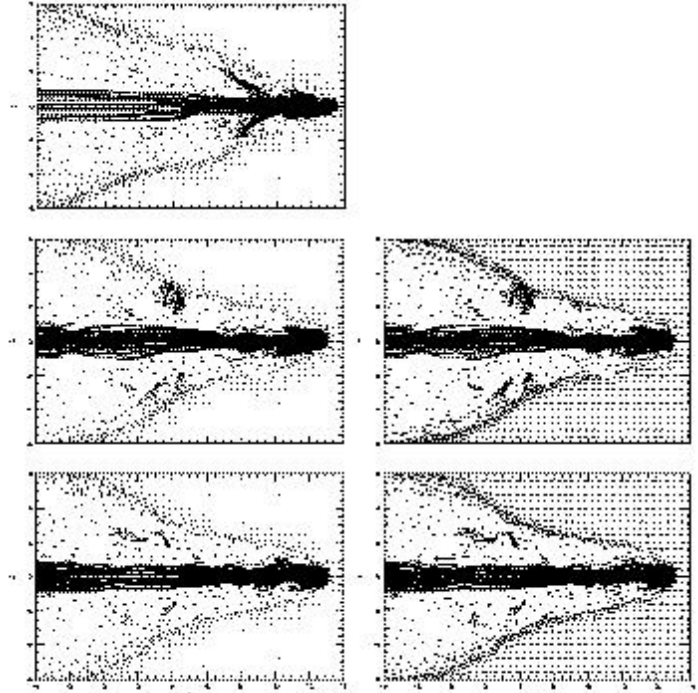

**Figures 2 and 3:** Mid-plane velocity field (left) and magnetic field (right) distributions of the jets of **Fig.1**: **HD3r** (top), **ML3r** (middle), and **MH3r** (bottom).